\documentclass[twocolumn,showpacs,prb]{revtex4}
\usepackage{amssymb}
\usepackage{makeidx}
\usepackage{amsmath}
\usepackage{graphicx}
\usepackage{dcolumn}
\usepackage{bm}
\usepackage{epstopdf}

\begin{document}

\title{Collective two-boson decay of excitons in Bose-Einstein condensate
and generation of coherent photon-phonon radiation}
\author{H.K. Avetissian}
\author{A.K. Avetissian}
\author{G.F. Mkrtchian}
\author{B.R. Avchyan}
\affiliation{Centre of Strong Fields Physics, Yerevan State University, 1 A. Manukian,
Yerevan 0025, Armenia}

\begin{abstract}
The collective decay of excitons from initial Bose-Einstein condensate state
is investigated theoretically. As practically more interesting case we
consider excitons of the yellow series in the semiconductor cuprous oxide
where we have collective photon and phonon assisted decay of excitons. It is
shown that because of intrinsic instability of recoilless two-boson decay of
Bose-Einstein condensate, the spontaneously emitted bosonic pairs are
amplified leading to an exponential buildup of a macroscopic population into
the certain modes. The collective decay rate has a nonlinear dependence on
the excitonic density being comparable or larger than Auger recombination
loss rate up to the high densities, which makes obtainable its observation.
The considering phenomenon can also be applied for the realization of phonon
laser.
\end{abstract}

\maketitle



\section{Introduction}

Over the past half-century, excitons were considered as notably interesting
candidates for Bose--Einstein condensation (BEC), in which collective
coherence may lead to intriguing macroscopic quantum phenomena (see, Ref. [%
\onlinecite{Moskal}] and references therein). Exciton being a bound state of
an electron and a hole in a semiconductor is a unique physical system with a
rather small mass comparable to the free electron mass. This is a crucial
advantage from the experimental point of view since the BEC critical
temperature of an excitonic gas is much higher than that of an atom gas with
the same number density. \cite{Pethick} However, the BEC was first
successfully realized for trapped alkali atoms, \cite{BEC} which are several
thousand times heavier than excitons. The latter provided additional
stimulus for realization of BEC for various condensed matter physics of
quasiparticles. In this context, it is worthy to mention realization of BEC
of quasiparticles, known as exciton--polaritons, \cite{Kasp} existing even
at room-temperature. \cite{BECS}

Among the variety of bosonic quasiparticles, the excitons of the yellow
series in the semiconductor cuprous oxide ($\mathrm{Cu}_{2}\mathrm{O}$) are
still considered as the most promising candidates for pure\ excitonic BEC. 
\cite{Rev1,Rev2,Rev3} Experiments in this direction have been done since
1986, \cite{Exp1,Exp2,Exp3,Exp4,Exp5} and continued up to present \cite%
{R1,R2,R3,R4,R5,R6,R7,R8} due to several favorable features of excitons in $%
\mathrm{Cu}_{2}\mathrm{O}$. First, the large binding energy of $0.15\ 
\mathrm{eV}$ which increases the Mott density up to $10^{19}\mathrm{cm}^{-3}$%
. Second, the ground state of this series splits into the threefold
degenerate orthoexciton and the non-degenerate paraexciton. The latter is
the lowest energetic state lying below the orthoexciton states. Due to the
selection rules one photon decay of paraexciton is forbidden. Its decay is
only possible via optical phonon and photon resulting in a long lifetime. 
\cite{Mys,Shi} The latter is in the microsecond range during which BEC may
be reached. To achieve excitonic BEC one should create a dense gas of
excitons either in a bulk crystal or in a potential trap. However
experiments \cite{Exp1,Exp2,Exp3,Exp4,Exp5,R3,R4,R5,R6} did not demonstrate
conclusively excitonic BEC. The main reason for this failure is connected
with the fact that the lifetime of excitons in $\mathrm{Cu}_{2}\mathrm{O}$
decreases significantly at high gas densities. This effect has been
attributed to an Auger recombination process between two excitons resulting
in a loss rate $\Gamma _{\mathrm{A}}=\alpha n$, where $\alpha $ is the Auger
constant and $n$ is the exciton gas density. However, there is no general
consent on the value of Auger constant. The reported values for $\alpha $
range are from $10^{-20}\ \mathrm{cm}^{3}\mathrm{ns}^{-1}$ to $1.8\times
10^{-16}\ \mathrm{cm}^{3}\mathrm{ns}^{-1}$ and differ for orto- and
para-excitons. \cite{D1,D2,D3,D4,D5,D6}

As was mentioned above, the isolated exciton in $\mathrm{Cu}_{2}\mathrm{O}$
is unstable and decays into photon and phonon. Due to BEC coherence, one can
expect collective radiative effects at the decay of a large number of
excitons. The latter may be a tool that evidences the state of the BEC, as
well as, it may significantly reduce the lifetime of the BEC state. Such an
effect has been revealed for the positronium atoms, \cite{Mer1,Mer2} which
in some sense resembles excitons. It has been shown that at the coupling of
two coherent ensembles of bosons -- the BEC of positronium atoms and photons
there is an instability at which, starting from the vacuum state of the
photonic field, the expectation value of the photon's mode occupation grows
exponentially for a narrow interval of frequencies. For the excitons in $%
\mathrm{Cu}_{2}\mathrm{O}$ one will have coupling between three bosonic
fields and it is of interest to investigate how excitonic BEC burst into
photons/phonons.

In this paper collective decay of excitons from initial Bose-Einstein
condensate state is investigated arising from the second quantized
formalism. It is shown that because of intrinsic instability of recoilless
two-boson decay of Bose-Einstein condensate, the spontaneously emitted
bosonic pairs are amplified, leading to an exponential buildup of a
macroscopic population into certain modes. The exponential growth rate has a
nonlinear dependence on the BEC density and it is quite large for the
experimentally achievable densities. For the elongated condensate, one can
reach self-amplification of the end-fire-modes. With the initial
monochromatic photonic beam, one can generate the monochromatic phononic
beam. Hence, the considered phenomenon may also be applied for realization
of phonon laser.

The paper is organized as follows. In Sec. II the main Hamiltonian is
introduced. In Sec. III spontaneous decay of exciton is analyzed. In Sec. IV
we consider intrinsic instability of recoilless collective two-boson decay
of excitonic BEC. Finally, conclusions are given in Sec. V.

\section{Basic Hamiltonian}

We start our study with the construction of the Hamiltonian which governs
the quantum dynamics of considered process. The total Hamiltonian consists
of four parts: 
\begin{equation}
\widehat{H}=\widehat{H}_{\mathrm{exc}}+\widehat{H}_{\mathrm{phot}}+\widehat{H%
}_{\mathrm{phon}}+\widehat{H}_{\mathrm{d}}.  \label{fH}
\end{equation}%
Here the first part is the Hamiltonian of free excitons:%
\begin{equation}
\widehat{H}_{\mathrm{exc}}=\int d\Phi _{\mathbf{p}}\mathcal{E}_{\text{%
\textsc{e}}}\left( \mathbf{p}\right) \widehat{\text{\textsc{e}}}_{\mathbf{p}%
}^{+}\widehat{\text{\textsc{e}}}_{\mathbf{p}},  \label{H_exc}
\end{equation}%
where $\widehat{\text{\textsc{e}}}_{\mathbf{p}}^{+}$ ($\widehat{\text{%
\textsc{e}}}_{\mathbf{p}}$) is the creation (annihilation) operator for an
exciton. These operators satisfy the Bosonic commutation rules for a
relatively small number density $n$ of excitons, that is at $n<n_{M}$, where 
$n_{M}$ is the Mott density. \cite{Moskal} For the integration in
phase-space we have introduced the notation $d\Phi _{\mathbf{p}}=\mathcal{V}%
d^{3}\mathbf{p}/\left( 2\pi \right) ^{3}$ ($\mathcal{V}$ is the quantization
volume). Then, $\mathcal{E}_{\text{\textsc{e}}}\left( \mathbf{p}\right)
=\hbar ^{2}\mathbf{p}^{2}/2m_{\ast }+\hbar \omega _{\mathrm{exc}}$ is the
total energy of exciton with the momentum $\hbar \mathbf{p}$ of the
center-of-mass motion, \textrm{\ }$m_{\ast }$ is an exciton mass, $\mathcal{E%
}_{in}$ is the exciton internal energy ($\hbar \omega _{\mathrm{exc}}=%
\mathcal{E}_{G}-\mathcal{E}_{b}$, in terms of the band-gap difference $%
\mathcal{E}_{G}$ and the binding energy $\mathcal{E}_{b}$).

\bigskip The second term in Eq. (\ref{fH}) is the Hamiltonian of the free
photons%
\begin{equation}
\widehat{H}_{\mathrm{phot}}=\int d\Phi _{\mathbf{k}}\hbar \omega \left( 
\mathbf{k}\right) \widehat{c}_{\mathbf{k}}^{+}\widehat{c}_{\mathbf{k}},
\label{H_ph}
\end{equation}%
where $\widehat{c}_{\mathbf{k}}$ ($\widehat{c}_{\mathbf{k}}^{+}$) is the
annihilation (creation) operator of the photon with the momentum $\mathbf{k}$
and dispersion relation $\omega =\omega \left( \mathbf{k}\right) $\textbf{. }%
The third term in Eq. (\ref{fH}) is the Hamiltonian of the free phonons with
annihilation (creation) operator $\widehat{b}_{\mathbf{q}}$ ($\widehat{b}_{%
\mathbf{q}}^{+}$): 
\begin{equation}
\widehat{H}_{\mathrm{phon}}=\int d\Phi _{\mathbf{q}}\hbar \omega _{\mathrm{ph%
}}\left( \mathbf{q}\right) \widehat{b}_{\mathbf{q}}^{+}\widehat{b}_{\mathbf{q%
}}.  \label{H_op}
\end{equation}%
The last term in Eq. (\ref{fH})%
\begin{equation*}
\widehat{H}_{\mathrm{d}}=\int d\Phi _{\mathbf{q}}\int d\Phi _{\mathbf{p}}%
\left[ \frac{\hbar \mathcal{M}\left( \mathbf{q},\mathbf{p}\right) }{\mathcal{%
V}^{1/2}}\widehat{b}_{\mathbf{q}}^{+}\widehat{c}_{\mathbf{p-q}}^{+}\widehat{%
\text{\textsc{e}}}_{\mathbf{p}}\right] 
\end{equation*}%
\begin{equation}
+\left. \frac{\hbar \mathcal{M}^{\ast }\left( \mathbf{q},\mathbf{p}\right) }{%
\mathcal{V}^{1/2}}\widehat{\text{\textsc{e}}}_{\mathbf{p}}^{+}\widehat{c}_{%
\mathbf{p-q}}\widehat{b}_{\mathbf{q}}\right]   \label{H_2j}
\end{equation}%
is the Hamiltonian of the two-boson decay of an exciton. \cite{Shi} Here we
assume that the direct recombination of electrons and holes is very weak and
the main decay process is a phonon-assisted recombination process in which
an exciton decays emitting an optical phonon, as well as, a photon. The
amplitude $\mathcal{M}\left( \mathbf{q},\mathbf{p}\right) $ for an exciton
decay can be calculated by the Feynman diagrams.

\section{Spontaneous decay of an exciton}

Before considering collective decay of excitons it will be useful to
consider spontaneous decay of a single exciton from the quantum dynamic
point of view. For the spontaneous decay we consider initial condition in
which the photonic and phononic fields begin in the vacuum state, while
excitonic field is prepared in a Fock state with the one exciton in the rest
($\mathbf{p=0}$). Such state can be represented as $|\Psi \left( 0\right)
\rangle =|0\rangle _{\mathrm{phon}}\otimes |0\rangle _{\mathrm{phot}}\otimes 
\widehat{\text{\textsc{e}}}_{0}^{+}|0\rangle _{\mathrm{exc}}$. Then the
state vector for times $t>0$ is just given by the expansion%
\begin{equation*}
|\Psi \rangle =C_{0}e^{-\frac{i}{\hbar }\mathcal{E}_{\text{\textsc{e}}%
}\left( 0\right) t}|0\rangle _{\mathrm{phon}}\otimes |0\rangle _{\mathrm{phot%
}}\otimes \widehat{\text{\textsc{e}}}_{0}^{+}|0\rangle _{\mathrm{exc}}+\int
d\Phi _{\mathbf{k}}d\Phi _{\mathbf{k}^{\prime }}
\end{equation*}%
\begin{equation}
\times C_{\mathbf{k};\mathbf{k}^{\prime }}\left( t\right) e^{-i\left( \omega
_{\mathrm{ph}}\left( \mathbf{k}^{\prime }\right) +\omega \left( \mathbf{k}%
\right) \right) t}\widehat{b}_{\mathbf{k}^{\prime }}^{+}|0\rangle _{\mathrm{%
phon}}\otimes \widehat{c}_{\mathbf{k}}^{+}|0\rangle _{\mathrm{phot}}\otimes
|0\rangle _{\mathrm{exc}},  \label{Psit}
\end{equation}%
where $C_{\mathbf{k};\mathbf{k}^{\prime }}\left( t\right) $ is the
probability amplitude for the photonic and phononic fields to be in the
single-particle state, while excitonic field in the vacuum state. From the
Schr\"{o}dinger equation one can obtain evolution equations:%
\begin{equation*}
i\frac{\partial C_{\mathbf{k};\mathbf{k}^{\prime }}\left( t\right) }{%
\partial t}=\frac{\mathcal{M}\left( \mathbf{k}^{\prime },\mathbf{0}\right) }{%
\mathcal{V}^{1/2}}C_{0}\left( t\right) \frac{\left( 2\pi \right) ^{3}}{%
\mathcal{V}}\delta \left( \mathbf{k}+\mathbf{k}^{\prime }\right) 
\end{equation*}%
\begin{equation}
\times e^{i\left( \omega _{\mathrm{ph}}\left( \mathbf{k}^{\prime }\right)
+\omega \left( \mathbf{k}\right) -\omega _{\mathrm{exc}}\right) t}.
\label{ev1}
\end{equation}%
Then, according to perturbation theory we take $C_{0}\left( t\right) \simeq 1
$, and for the amplitude $C_{\mathbf{k};\mathbf{k}^{\prime }}\left(
t\rightarrow \infty \right) $ from Eq. (\ref{ev1}) we obtain%
\begin{equation*}
C_{\mathbf{k};\mathbf{k}^{\prime }}=\frac{\mathcal{M}\left( \mathbf{k}%
^{\prime },\mathbf{0}\right) }{i\mathcal{V}^{1/2}}\frac{\left( 2\pi \right)
^{4}}{\mathcal{V}}
\end{equation*}%
\begin{equation}
\times \delta \left( \omega _{\mathrm{ph}}\left( \mathbf{k}^{\prime }\right)
+\omega \left( \mathbf{k}\right) -\omega _{\mathrm{exc}}\right) \delta
\left( \mathbf{k}+\mathbf{k}^{\prime }\right) .  \label{pert3}
\end{equation}%
For the decay of an exciton the modes laying in the narrow interval of
wavenumbers are responsible. Hence, for the dispersion relations we assume $%
\omega _{\mathrm{ph}}\left( \mathbf{k}\right) =\mathrm{const}\equiv \omega _{%
\mathrm{ph}}$ and $\omega \left( \mathbf{k}\right) =kc_{l}$, where $c_{l}$%
-is the light speed in a semiconductor. Then returning to expansion (\ref%
{Psit}), one can write%
\begin{equation*}
|\Psi \rangle \simeq C_{0}e^{-i\omega _{\mathrm{exc}}t}|0\rangle _{\mathrm{%
phon}}\otimes |0\rangle _{\mathrm{phot}}\otimes |1_{\mathbf{0}}\rangle _{%
\mathrm{exc}}
\end{equation*}%
\begin{equation*}
+\frac{\mathcal{V}^{1/2}\mathcal{M}\left( k_{0},0\right) k_{0}^{2}}{i\left(
2\pi \right) ^{2}c_{l}}e^{-i\omega _{\mathrm{exc}}t}
\end{equation*}%
\begin{equation}
\times |0\rangle _{\mathrm{exc}}\otimes \int d\widehat{\mathbf{k}}|1_{%
\mathbf{k}}\rangle _{\mathrm{phon}}\otimes |1_{-\mathbf{k}}\rangle _{\mathrm{%
phot}},  \label{fs}
\end{equation}%
\ where $\widehat{\mathbf{k}}=\mathbf{k/}\left\vert \mathbf{k}\right\vert $,
and 
\begin{equation}
k_{0}=\frac{\omega _{\mathrm{exc}}-\omega _{\mathrm{ph}}}{c_{l}}.  \label{k0}
\end{equation}%
Hear we have taken into account that the decay amplitude does not depend on
the direction of $\mathbf{k}$ and\textbf{, }as a result, the final state (%
\ref{fs}) resulting from an exciton decay is a superposition of the states
of oppositely propagating photon and phonon with the given momentum $k_{0}$.
That is, we have recoilless two-boson decay of exciton, which is crucial for
the development of instability in BEC where the lowest energy single
particle state is occupied. For the decay rate one can write 
\begin{equation*}
\Gamma =\int d\Phi _{\mathbf{k}}d\Phi _{\mathbf{k}^{\prime }}\frac{%
\left\vert C_{\mathbf{k};\mathbf{k}^{\prime }}\right\vert ^{2}}{t_{\mathrm{%
int}}},
\end{equation*}%
where $t_{\mathrm{int}}$ is the interaction time. With the help of Eq. (\ref%
{pert3}) we obtain the well known result: 
\begin{equation}
\Gamma =\frac{\mathcal{M}^{2}}{\pi c_{l}}k_{0}^{2}.  \label{456}
\end{equation}%
The radiative lifetime of an isolated exciton is $\Gamma ^{-1}$.

\section{Collective decay}

For analysis of the collective photon-phonon decay of excitons we will use
Heisenberg representation, where the evolution operators are given by the
following equation 
\begin{equation}
i\frac{\partial \widehat{L}}{\partial t}=\left[ \widehat{L},\widehat{H}%
\right] ,  \label{Heis}
\end{equation}%
and the expectation values are determined by the initial wave function $\Psi
_{0}$:%
\begin{equation*}
\left\langle \widehat{L}\right\rangle =\langle \Psi _{0}|\widehat{L}|\Psi
_{0}\rangle .
\end{equation*}%
We will assume that the excitonic field starts up in the Bose-Einstein
condensate state, while for photonic and phononic fields we will consider
both vacuum state and states with nonzero mean number of particles. Taking
into account Hamiltonian (\ref{fH}) from Eq. (\ref{Heis}) we obtain a set of
equations:%
\begin{equation}
i\frac{\partial \widehat{c}_{\mathbf{k}}}{\partial t}=\omega \left( \mathbf{k%
}\right) \widehat{c}_{\mathbf{k}}+\int d\Phi _{\mathbf{p}}\frac{\mathcal{M}%
\left( \mathbf{p-k},\mathbf{p}\right) }{\mathcal{V}^{1/2}}\widehat{b}_{%
\mathbf{p-k}}^{+}\widehat{\text{\textsc{e}}}_{\mathbf{p}},  \label{M1}
\end{equation}%
\begin{equation}
i\frac{\partial \widehat{b}_{\mathbf{k}}}{\partial t}=\omega _{\mathrm{ph}%
}\left( \mathbf{k}\right) \widehat{b}_{\mathbf{k}}+\int d\Phi _{\mathbf{p}}%
\frac{\mathcal{M}\left( \mathbf{k},\mathbf{p}\right) }{\mathcal{V}^{1/2}}%
\widehat{c}_{\mathbf{p-k}}^{+}\widehat{\text{\textsc{e}}}_{\mathbf{p}},
\label{M2}
\end{equation}%
\begin{equation}
i\frac{\partial \widehat{\text{\textsc{e}}}_{\mathbf{p}}}{\partial t}=\hbar
^{-1}\mathcal{E}_{\text{\textsc{e}}}\left( \mathbf{p}\right) \widehat{\text{%
\textsc{e}}}_{\mathbf{p}}+\int d\Phi _{\mathbf{q}}\frac{\mathcal{M}^{\ast
}\left( \mathbf{q},\mathbf{p}\right) }{\mathcal{V}^{1/2}}\widehat{c}_{%
\mathbf{p-q}}\widehat{b}_{\mathbf{q}}.  \label{M3}
\end{equation}

These equations are a nonlinear set of equations with the photonic, phononic
and excitonic fields' operators defined self-consistently. As we are
interested in the quantum dynamics of considered system in the presence of
instabilities we can decouple the excitonic field treating the dynamics of
photonic and phononic fields. For this propose we just use the Bogolubov
approximation. If the lowest energy single particle state has a macroscopic
occupation, we can separate the field operators $\widehat{\text{\textsc{e}}}%
_{\mathbf{p}}$ into the condensate term and the non-condensate components,
i.e. the operator $\widehat{\text{\textsc{e}}}_{\mathbf{p}}$ in Eqs. (\ref%
{M1}) and (\ref{M2}) is replaced by the c-number as follow%
\begin{equation}
\widehat{\text{\textsc{e}}}_{\mathbf{p}}=\sqrt{n_{0}}\frac{\left( 2\pi
\right) ^{3}}{\mathcal{V}^{1/2}}\delta \left( \mathbf{p}\right) e^{-i\omega
_{\mathrm{exc}}t},  \label{Bogol}
\end{equation}%
where $n_{0}$ is the number density of excitons in the condensate. Making
Bogoloubov approximation we arrive at a linear set of the Heisenberg
equations%
\begin{equation}
i\frac{\partial \widehat{c}_{-\mathbf{k}}}{\partial t}=\omega \left(
k\right) \widehat{c}_{-\mathbf{k}}+\chi \left( k\right) \widehat{b}_{\mathbf{%
k}}^{+}e^{-i\omega _{\mathrm{exc}}t},  \label{L1}
\end{equation}%
\begin{equation}
i\frac{\partial \widehat{b}_{\mathbf{k}}}{\partial t}=\omega _{\mathrm{ph}}%
\widehat{b}_{\mathbf{k}}+\chi \left( k\right) \widehat{c}_{\mathbf{-k}%
}^{+}e^{-i\omega _{\mathrm{exc}}t},  \label{L2}
\end{equation}%
which couples photon modes with momentum $\mathbf{k}$ to the phonons with
momentum $-\mathbf{k}$. The coupling constant is 
\begin{equation}
\chi \left( k\right) =\sqrt{n_{0}}\mathcal{M}\left( k,0\right) .  \label{CC}
\end{equation}

Equations (\ref{L1}) and (\ref{L2}) compose a set of linearly coupled
operator equations that can be solved by the method of characteristics whose
eigenfrequencies define the temporal dynamics of the bosonic fields. The
existence of an eigenfrequency with an imaginary part would indicate the
onset of instability at which the initial spontaneously emitted bosonic
pairs are amplified leading to an exponential buildup of a macroscopic mode
population. Solving Eqs. (\ref{L1}) and (\ref{L2}), we obtain%
\begin{equation*}
\widehat{b}_{\mathbf{k}}^{+}=e^{i\left( \omega _{\mathrm{ph}}-\frac{\delta
\left( k\right) }{2}\right) t}\left\{ \widehat{b}_{\mathbf{k}}^{+}\left(
0\right) \cosh \left( \sigma \left( k\right) t\right) +\frac{i}{2\sigma
\left( k\right) }\right.
\end{equation*}%
\begin{equation}
\left. \times \left( \delta \left( k\right) \widehat{b}_{\mathbf{k}%
}^{+}\left( 0\right) +2\chi ^{\ast }\left( k\right) \widehat{c}_{-\mathbf{k}%
}\left( 0\right) \right) \sinh \left( \sigma \left( k\right) t\right)
\right\} ,  \label{S1}
\end{equation}%
\begin{equation*}
\widehat{c}_{-\mathbf{k}}=e^{i\left( \frac{\delta \left( k\right) }{2}%
-\omega \left( k\right) \right) t}\left\{ \widehat{c}_{-\mathbf{k}}\left(
0\right) \cosh \left( \sigma \left( k\right) t\right) -\frac{i}{2\sigma
\left( k\right) }\right.
\end{equation*}%
\begin{equation}
\left. \times \left( 2\chi \left( k\right) \widehat{b}_{\mathbf{k}%
}^{+}\left( 0\right) +\delta \left( k\right) \widehat{c}_{-\mathbf{k}}\left(
0\right) \right) \sinh \left( \sigma \left( k\right) t\right) \right\} ,
\label{S2}
\end{equation}%
where%
\begin{equation}
\delta \left( k\right) =\omega \left( k\right) -\omega _{\mathrm{exc}%
}+\omega _{\mathrm{ph}}  \label{det}
\end{equation}%
is the resonance detuning, and 
\begin{equation}
\sigma \left( k\right) =\sqrt{\left\vert \chi \left( k\right) \right\vert
^{2}-\frac{\delta ^{2}\left( k\right) }{4}}.  \label{CCC}
\end{equation}%
As is seen from Eqs. (\ref{S1})-(\ref{CCC}), the condition for the dynamic
instability is: 
\begin{equation*}
\left\vert \chi \left( k\right) \right\vert >\frac{\left\vert \delta \left(
k\right) \right\vert }{2}
\end{equation*}%
leading to the exponential growth of the modes in the narrow interval of
wavenumbers%
\begin{equation}
\omega _{\mathrm{exc}}-\omega _{\mathrm{ph}}-2\left\vert \chi \left(
k_{0}\right) \right\vert <kc_{l}<\omega _{\mathrm{exc}}-\omega _{\mathrm{ph}%
}+2\left\vert \chi \left( k_{0}\right) \right\vert .  \label{inter}
\end{equation}%
For the interval (\ref{inter}) we find that the expectation value of the
photonic and phononic modes occupations grow exponentially:%
\begin{equation*}
N_{\mathrm{phot}}\left( \mathbf{k,}t\right) =\langle \Psi _{0}|\widehat{c}_{%
\mathbf{k}}^{+}\widehat{c}_{\mathbf{k}}|\Psi _{0}\rangle
\end{equation*}%
\begin{equation*}
=N_{\mathrm{phot}}\left( \mathbf{k,}0\right) \left( \cosh ^{2}\left( \sigma
\left( k\right) t\right) +\frac{\delta ^{2}\left( k\right) }{4\sigma
^{2}\left( k\right) }\sinh ^{2}\left( \sigma \left( k\right) t\right) \right)
\end{equation*}%
\begin{equation}
+\frac{\left\vert \chi \left( k\right) \right\vert ^{2}}{\sigma ^{2}\left(
k\right) }\left( 1+N_{\mathrm{phon}}\left( -\mathbf{k,}0\right) \right)
\sinh ^{2}\left( \sigma \left( k\right) t\right) ,  \label{Exp1}
\end{equation}%
\begin{equation*}
N_{\mathrm{phon}}\left( \mathbf{k,}t\right) =\langle \Psi _{0}|\widehat{b}_{%
\mathbf{k}}^{+}\widehat{b}_{\mathbf{k}}|\Psi _{0}\rangle
\end{equation*}%
\begin{equation*}
=N_{\mathrm{phon}}\left( \mathbf{k,}0\right) \left( \cosh ^{2}\left( \sigma
\left( k\right) t\right) +\frac{\delta ^{2}\left( k\right) }{4\sigma
^{2}\left( k\right) }\sinh ^{2}\left( \sigma \left( k\right) t\right) \right)
\end{equation*}%
\begin{equation}
+\frac{\left\vert \chi \left( k\right) \right\vert ^{2}}{\sigma ^{2}\left(
k\right) }\left( 1+N_{\mathrm{phot}}\left( -\mathbf{k,}0\right) \right)
\sinh ^{2}\left( \sigma \left( k\right) t\right) .  \label{Exp2}
\end{equation}%
For the central wavenumber ($\delta \left( k_{0}\right) =0$) the exponential
growth rate is 
\begin{equation}
G=2\chi \left( k_{0}\right) =2\sqrt{n_{0}}\mathcal{M}\left( k_{0},0\right) .
\label{gain1}
\end{equation}%
Taking into account Eq. (\ref{CC}) and derived expression (\ref{456}) for
the decay rate, we obtain compact expression for the exponential growth rate:%
\begin{equation}
G=\sqrt{\frac{4\pi n_{0}c_{l}\Gamma }{k_{0}^{2}}}.  \label{main}
\end{equation}

As is seen from Eqs. (\ref{Exp1}) and (\ref{Exp2}), we have an exponential
buildup of a macroscopic mode population even for the initial vacuum state $%
N_{\mathrm{phot}}\left( \mathbf{k,}0\right) =N_{\mathrm{phon}}\left( \mathbf{%
k,}0\right) =0$. In this case from Eqs. (\ref{Exp1}) and (\ref{Exp2}) we have%
\begin{equation*}
N_{\mathrm{phot}}\left( \mathbf{k,}t\right) =N_{\mathrm{phon}}\left( \mathbf{%
k,}t\right) =\frac{4\left\vert \chi \left( k\right) \right\vert ^{2}}{%
4\left\vert \chi \left( k\right) \right\vert ^{2}-\delta ^{2}\left( k\right) 
}
\end{equation*}%
\begin{equation}
\times \left( e^{\sqrt{4\left\vert \chi \left( k\right) \right\vert
^{2}-\delta ^{2}\left( k\right) }t}+e^{-\sqrt{4\left\vert \chi \left(
k\right) \right\vert ^{2}-\delta ^{2}\left( k\right) }t}-2\right) .
\label{izo}
\end{equation}

We have solved the issue considering uniform BEC without boundary conditions
and, as a consequence, according to Eq. (\ref{izo}) we have an isotropic
exponential gain. Due to the BEC coherence, here we have an absolute
instability, i.e., the number of photons/phonons grows at every point within
a BEC and the gain is scaled as $\sqrt{n_{0}}$. Here the excitonic BEC burst
into photons and phonons. Note, that our approximation is valid for the
interaction times $t_{\mathrm{int}}$ at which the total number of photons
and phonons are much smaller than the number of excitons in BEC: $N_{\mathrm{%
phot}},N_{\mathrm{phon}}<<N_{\mathrm{exc}}$.

For laserlike action, i.e., for directional radiation, one should take an
elongated shape of the BEC. In this case, boundary conditions define
interaction time. This can be incorporated into the derived equation (\ref%
{L1}) and (\ref{L2}) by introducing mode damping. The latter is simply due
to the propagation of the bosonic fields, which escapes from the active
medium and is inversely proportional to the transit time of a photon in the
active medium. This transit time strictly depends on the propagation
direction. The latter is equivalent to the finite interaction time strictly
depending on the shape of the BEC.

For the directional radiation decay, one can also consider initial photonic
or phononic beam. For the initial monochromatic photonic beam, in the result
of the collective decay, one will have backscattered monochromatic phononic
beam. Thus, one can realize a coherent source of phonons applying resonant
laser beam.

Let us make explicit calculations for the initial photonic beam with the
distribution: 
\begin{equation*}
N_{\mathrm{phot}}\left( \mathbf{k,}0\right) =N_{0}\exp \left( -\frac{%
k_{x}^{2}+k_{y}^{2}}{2\delta ^{2}}\right) \exp \left( -\frac{\left(
k_{z}-k_{0}\right) ^{2}}{2\delta ^{2}}\right) ,
\end{equation*}%
where $N_{0}>>1$ and $\delta $ is the width of distribution in the momentum
space $\delta <<k_{0}$. In this case, for the angular distribution of the
phonon number density (we assume $N_{\mathrm{phon}}\left( \mathbf{k,}%
0\right) <<1$) we have%
\begin{equation*}
\frac{dn_{\mathrm{phon}}}{d\vartheta }\simeq \frac{N_{0}}{\left( 2\pi
\right) ^{2}}\int_{k_{0}-\frac{G}{c_{l}}}^{k_{0}+\frac{G}{c_{l}}}dk\frac{%
k^{2}\left\vert \chi \left( k\right) \right\vert ^{2}}{\sigma ^{2}\left(
k\right) }\sin \vartheta
\end{equation*}%
\begin{equation}
\times \exp \left( -\frac{k^{2}\sin ^{2}\vartheta }{2\delta ^{2}}-\frac{%
\left( k\cos \vartheta +k_{0}\right) ^{2}}{2\delta ^{2}}\right) \sinh
^{2}\left( \sigma \left( k\right) t_{\mathrm{int}}\right) ,  \label{nphon}
\end{equation}%
where $t_{\mathrm{int}}$ is the interaction time of the photonic beam with
excitonic BEC. As is seen from Eq. (\ref{nphon}), phonons are radiated in
the opposite to the photonic beam direction and have peak near $\vartheta
\simeq \pi $. For the phonon number density one should integrate Eq. (\ref%
{nphon}) over $\vartheta $. Taking into account that $G<<k_{0}c_{l}$, we
obtain: 
\begin{equation}
n_{\mathrm{phon}}\simeq \frac{\delta ^{2}GN_{0}}{2\pi ^{2}c_{l}}F\left( Gt_{%
\mathrm{int}}\right) ,  \label{total}
\end{equation}%
where 
\begin{equation}
F\left( Gt_{\mathrm{int}}\right) =\int_{0}^{1}dx\frac{\sinh ^{2}\left( \frac{%
Gt_{\mathrm{int}}}{2}\sqrt{1-x^{2}}\right) }{1-x^{2}}  \label{fanct}
\end{equation}%
is the amplification factor. The latter is a rapidly increasing function,
displayed in\ Fig. 1. 
\begin{figure}[tbp]
\includegraphics[width=.50\textwidth]{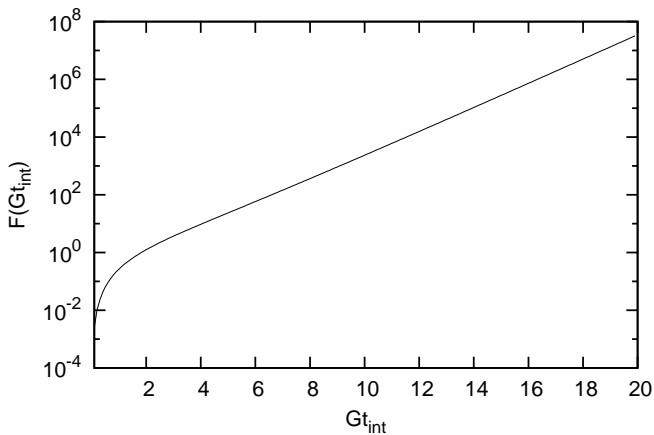}
\caption{In the logarithmic scale it is shown the dimensionless
amplification factor versus interaction time in units of $G^{-1}$.}
\end{figure}

Let us consider the parameters required for observation of the considered
effect for the excitons of the yellow series in the semiconductor cuprous
oxide. In Cu$_{2}$O, the radiative lifetime of an isolated exciton is $%
\Gamma ^{-1}\approx 10^{-5}\mathrm{s}$, the refractive index is
approximately $3$ ($c_{l}\approx 10^{10}\mathrm{cm/s}$), the energy of the
optical phonon is $10^{-2}$ $\mathrm{eV}$, the energy gap $\mathcal{E}%
_{G}\approx 2$ $\mathrm{eV}$ and the binding energy $\mathcal{E}_{b}\approx
0.15$ $\mathrm{eV}$. Thus,\textrm{\ }for the exponential growth rate we have:%
\begin{equation}
G\simeq \left( \frac{n_{0}}{10^{18}\mathrm{cm}^{-3}}\right) ^{1/2}\times
4\times 10^{11}\mathrm{s}^{-1}.  \label{Cu2O}
\end{equation}%
As is seen from Eq. (\ref{Cu2O}), the growth rate is quite large $G\simeq
4\times 10^{11}\mathrm{s}^{-1}$for the experimentally achievable densities $%
n_{0}=10^{18}\mathrm{cm}^{-3}$. Note that collective growth rate is larger
than Auger recombination loss rate $\Gamma _{\mathrm{A}}=\alpha n_{0}$ up to
high densities $n_{0}<4\times 10^{18}\mathrm{cm}^{-3}$. Let us also estimate
possible parameters of coherent phononic beam generated by the photon beam.
Taking density $n_{0}=10^{18}\mathrm{cm}^{-3}$ and interaction time $t_{%
\mathrm{int}}\approx 50\ \mathrm{ps\ }$from Fig.1 one can define $F\left(
Gt_{\mathrm{int}}\simeq 20\right) \simeq 3.5\times 10^{7}$. From Eq. (\ref%
{total}) for the phonon number density we have $n_{\mathrm{phon}}\simeq
N_{0}\times 5.6\times 10^{12}\mathrm{cm}^{-3}$. Thus, considered phenomenon
may be applied for the realization of a phonon laser.

\section{Conclusion}

In conclusion, we have studied the collective two-boson decay of excitons,
arising from the second quantized formalism. It was shown that BEC state is
unstable because of recoilless two-boson decay. The spontaneously emitted
bosonic pairs are amplified leading to an exponential buildup of a
macroscopic population into resonant modes. As a practically more
interesting case, we have considered the decay of excitons of the yellow
series in the semiconductor cuprous oxide, where BEC burst into the photons
and phonons with the collective growth rate proportional to the square root
of the BEC density. Calculations show that the collective decay rate is
comparable or larger than Auger recombination loss rate up to the high
densities. Hence, it can be used as a tool that evidences the formation of
BEC state in Cu$_{2}$O. We have also studied another application of
considered effect -- a possible source for generation of coherent phonon
beam. For the latter propose one can take an elongated condensate where
self-amplification of the end-fire-modes takes place. Otherwise, applying a
resonant photonic beam one can generate backscattered intense coherent
phonon beam.

\begin{acknowledgments}
This work was supported by the State Committee of Science MES RA, in the
frame of the research project SCS 15T-1C013.
\end{acknowledgments}

\end{document}